# Pseudogap in hole doped cuprates: a status report

*Saleh Hasan Naqib* and Raihana Shams Islam*

*Department of Physics, University of Rajshahi, Rajshahi 6205, Bangladesh*

*Corresponding author; Emails: jsr_eda1@ru.ac.bd (Saleh Hasan Naqib)

## ABSTRACT

The pseudogap state, an anomalous momentum dependent partial suppression of low-energy electronic excitations persisting well above the superconducting critical temperature ($T_c$) in underdoped cuprates, remains one of the most profound enigmas in condensed physics today. This review summarizes over three decades of experimental and theoretical progress. It appears that pseudogap is not a mere precursor to superconductivity, but a distinct, intertwined state of matter characterized by a complex interplay of competing and cooperating electronic orders. We critically examine the bulk probes like resistivity and magnetic susceptibility and the surface hallmark spectroscopic signatures observed via angle-resolved photoemission spectroscopy (ARPES) in this review. ARPES reveals a Fermi surface truncated into disconnected "arcs" and scanning tunneling microscopy (STM) visualizes a spatially heterogeneous energy gap around the Fermi level. These observations are contextualized alongside transport anomalies, including the in- and out-of-plane resistivity and bulk magnetic susceptibility, to construct a phenomenological portrait of a phase that violates conventional Fermi liquid theory. A central debate is the indications for underlying broken symmetries, from intra-unit-cell orbital loop currents detectable by spin-polarized neutron scattering to incipient charge density waves and short-range spin correlations. We evaluate leading theoretical frameworks - including performed Cooper pair models and scenarios of intertwined orders emanating from a quantum critical point deep within the superconducting dome at zero temperature. By mapping the temperature-doping phase diagram, we highlight the abrupt transition at above optimal doping. We conclude with open questions still to be answered. This review underscores the pseudogap's role as a key to unlocking the unconventional pairing mechanism of hole-doped cuprates and the broader physics of strongly correlated electron systems.

**Keywords:** Superconductivity in cuprates; Pseudogap; Quantum critical point; T-p phase diagram; Preformed pairs; Competing correlations

## 1. Introduction

The discovery of high-temperature superconductivity in doped copper oxide compounds (cuprates) marked a far-reaching departure from the conventional Bardeen-Cooper-Schrieffer (BCS) framework. Among the numerous anomalies found in the temperature versus hole content (T-p) phase diagram, none has proven more central or more widely contested than the pseudogap (PG). The strong electronic correlation signature was first characterized experimentally in underdoped cuprates via nuclear magnetic resonance (NMR) Knight-shift measurements and specific heat studies (Alloul et al., 1989; Loram et al., 1993; Tallon and Loram, 2001; Timusk and Statt, 1999; Loram et al., 1998a; Lee, 2008). Unlike the conventional pseudogap in the electronic energy density of states (EDOS) (Mott and Davis, 1979), pseudogap in high-$T_c$ cuprate superconductors refers to a prominent momentum dependent suppression in the low-energy electronic density of states which can make its presence felt at temperatures well above the superconducting transition temperature (Damascelli et al., 2003).

After four decades of discovery of superconductivity in hole doped copper oxide compounds (Bednorz and Müller, 1986), the underlying electronic mechanism responsible for Cooper pairing in these strongly correlated electronic materials is not a settled issue (Mueller, 2017; Tallon and Loram, 2001; Timusk and Statt, 1999; Loram et al., 1998a; Lee, 2008; Naqib et al., 2005a; Naqib et al., 2005b; Naqib et al., 2003). No definitive conclusions have been reached. It is widely believed that solving the pseudogap puzzle is the way forward to understand superconductivity in cuprates (Tallon and Loram, 2001; Timusk and Statt, 1999; Loram et al., 1998a; Lee, 2008; Naqib et al., 2005a; Naqib et al., 2005b; Naqib et al., 2003; Loram et al., 1998b).

The pseudogap in cuprates is characterized by a partial suppression of low-energy electronic spectral weight near the Fermi level, without the complete gap and long-range phase coherence associated with conventional superconductivity. PG was initially identified through nuclear magnetic resonance measurements, which revealed a reduction in the uniform spin susceptibility upon cooling in underdoped cuprates (Alloul et al., 1989; Naqib et al., 2009; Naqib and Islam, 2008). Spectroscopic probes - most notably Angle-Resolved Photoemission Spectroscopy (ARPES) and Scanning Tunneling Microscopy (STM) - reveal that the suppression of low-energy spectral weight preferentially occurs close to the anti-nodal regions of the two-dimensional Brillouin zone (BZ) suitable for the $CuO_2$ plane, while leaving gapless Fermi arcs intact along the diagonal nodal directions (Damascelli et al., 2003; Pan et al., 2000; Pan et al., 2001).

The PG is particularly prominent in the underdoped region (Bucher et al., 1993; Alloul et al., 1989), where its characteristic temperature, commonly denoted by $T^*$ largely exceeds $T_c$. Its evolution with temperature and carrier concentration provides essential information about the electronic structure and competing or intertwined interactions in these materials (Tallon and Loram, 2001; Timusk and Statt, 1999; Loram et al., 1998a; Lee, 2008; Naqib et al., 2005a; Naqib et al., 2005b; Naqib et al., 2003; Loram et al., 1998b). The PG exhibits a momentum dependence akin to the generic d-wave superconducting gap of cuprates, raising fundamental questions as to whether the two phenomena share a common microscopic origin or represent distinct electronic states. Although the pseudogap is often discussed in relation to superconductivity, its persistence above $T_c$ and its complex evolution across the phase diagram distinguish it from the conventional superconducting gap of BCS theory.

The physical origin of the PG remains highly controversial, with two broad classes of explanations receiving considerable attention (Tallon and Loram, 2001; Timusk and Statt, 1999; Loram et al., 1998a; Lee, 2008). In one, it is associated with precursor superconductivity: electron pairs form above $T_c$, but thermal or quantum phase fluctuations prevent the establishment of long-range Cooper pair phase coherence. This interpretation is motivated in part by the quasi-two-dimensional nature of cuprates and the possibility of a substantial separation between pair formation and phase ordering temperatures. In the other, the PG arises from an electronic instability distinct from superconducting pairing, potentially involving spin, charge or orbital ordering, or other forms of symmetry breaking (Tallon and Loram, 2001; Timusk and Statt, 1999; Loram et al., 1998a; Lee, 2008; Naqib et al., 2005a; Naqib et al., 2005b; Naqib et al., 2003; Loram et al., 1998b; Tallon et al., 1999). Experimental observations of charge-density-wave correlations and signatures of broken electronic symmetries have provided evidence relevant to these proposals, although their precise relationship to the PG is still not clear. Recent advancements in high-resolution spectroscopy, transport measurements, and quantum oscillations have added significant nuance to this binary framework, suggesting that strong Mott-Hubbard electron correlations and Mottness-induced electron fractionalization may intrinsically generate a pseudogap state from which both density waves and d-wave superconductivity emerge (Horio et al., 2025).

The relationship between the PG and superconductivity is therefore a central question in cuprate research. The PG may compete with superconductivity by removing low-energy electronic states that could otherwise participate in pairing, or it may be connected to the interactions that facilitate superconducting (SC) pairing. These possibilities are not necessarily mutually exclusive: the complex phase diagram may involve several intertwined electronic orders and fluctuating states whose relative importance changes with temperature, doping and material composition.

This review provides a critical overview of the PG phenomenon in hole doped cuprates. We examine the thermodynamic properties, trace its spectroscopic signatures across momentum space and critically examine the bulk transport and magnetic signatures (Loram et al., 2001; Loram et al., 1998b; Islam and Naqib, 2018; Naqib et al., 2016; Islam et al., 2010; Naqib et al., 2005a; Naqib et al., 2005b; Naqib et al., 2003). It is worth mentioning that PG is also observed in electron doped copper oxide superconductors (Armitage et al., 2010; Greene et al., 2020). There are strong similarities with hole doped systems but there are also distinct differences (Armitage et al., 2010; Greene et al., 2020; Weber et al., 2010). In this paper we focus on the PG observed in hole doped high-$T_c$ cuprates.

## 2. Experimental overview of the pseudogap

PG in cuprates was first identified experimentally in underdoped $YBa_2Cu_3O_{7-\delta}$ (Y123) through NMR measurements in 1989. $^{89}$Y Knight-shift measurements by Alloul, Ohno, and Mendels (Alloul et al., 1989) revealed a pronounced suppression of the uniform spin susceptibility well above $T_c$. This was interpreted as evidence for an incomplete gap in the low-energy electronic spin excitation spectrum as it is directly linked to the EDOS at the Fermi level.

The earliest convincing detection of the cuprate pseudogap through heat-capacity measurements is generally associated with the electronic specific-heat work of J. W. Loram and collaborators in the early 1990s (Loram et al., 1993), particularly their measurements on oxygen-doped Y123. The main point here was that the electronic entropy and specific heat evolved anomalously with underdoping. Loram and collaborators interpreted the systematic loss of low-energy electronic spectral weight as an evidence for a strong modification of the low-energy quasiparticle excitation spectrum (Loram et al., 1993).

Marshall et al. (1996) performed ARPES measurements on underdoped $Bi_2Sr_2CaCu_2O_{8+\delta}$ (Bi2212) and found clear evidence that sections of the Fermi surface disappeared in the normal state much above $T_c$. This observation was associated with an incomplete energy gap developing near $(\pi, 0)$ of the Brillouin zone above $T_c$. This was the first direct spectroscopic indications of the momentum dependent PG phenomenon. With momentum resolution, ARPES is an ideal technique to detect the low energy quasiparticle (QP) spectral weight of strongly correlated electronic materials.

Scanning tunneling microscopy and spectroscopy (STM/STS) have played a pivotal role in uncovering the spatial inhomogeneity and energy scale of the PG in cuprate high-temperature superconductors (Pan et al., 2001; Hoffman et al., 2002) and other correlated electron systems. STS/STM measures the real space differential conductance ($\frac{dI}{dV}$), which is proportional to the local density of states. In cuprates (such as $Bi_2Sr_2CaCu_2O_{8+\delta}$ or B2212), tunneling spectra reveal a suppression of QP spectral density around the Fermi level (Fischer et al., 2007; Pan et al., 2001). Above the superconducting critical temperature, this dip or "pseudogap" in the local electronic density of states persists up to a higher characteristic temperature ($T^*$), demonstrating that a partial energy gap exists in the normal state. Unlike conventional BCS superconductors, early and landmark STM studies revealed that the pseudogap and superconducting pairing gaps are inhomogeneous on the nanometer scale (Pan et al., 2001; Hoffman et al., 2002). The gap magnitude correlates strongly with local atomic environments (e.g., oxygen dopant locations), showing that electronic matter in the PG regime might be granular and strongly disordered on the nanoscale.

It should be noted that both ARPES and STS/STM are surface sensitive probes. On the other hand, resistivity, magnetic susceptibility and Hall effect measurements probe the bulk properties. PG has distinct effects on these bulk physical characteristics.

The PG in cuprate superconductors leads to a profound, highly anisotropic temperature dependent evolution of transport properties, fundamentally altering both in-plane ($\rho_{ab}$) and out-of-plane ($\rho_c$) resistivity. As the opening of the PG partially suppresses the electronic energy density of states at the Fermi level, leading to distinct transport signatures along different crystallographic axes due to the layered, quasi-two-dimensional nature of cuprates. In the normal state of optimally doped and overdoped cuprates, the in-plane resistivity exhibits a conventional linear temperature dependence over a wide temperature range - a hallmark of strange metal behavior. When the temperature is lowered below the pseudogap temperature ($T^*$), the depletion in the QP spectral weight effectively reduces the available charge carriers leading to the following behavior in $\rho_{ab}$ (T) (Naqib et al., 2005a; Naqib et al., 2005b; Naqib et al., 2003; Kokanović et al., 2006; Naqib et al., 2016): (i) As temperature decreases below a particular value $T^*$, $\rho_{ab}$ (T) deviates downward from the high-temperature linear trend, reflecting an overall loss of carrier scattering rate due to partial drop in the effective carrier density and (ii) in heavily underdoped cuprates, this suppression eventually leads to a broad maximum (or peak) in $\rho_{ab}$(T) at a temperature significantly lower than $T^*$, followed by a rapid decrease at lower temperatures as thermal scattering freezes out (prior to the onset of superconductivity).

Copper oxide superconductors are highly anisotropic. The conducting hole doped $CuO_2$ layers are separated by non-conducting charge reservoir segments making c-axis charge transport significantly weaker than that within the ab-plane (Naqib et al., 2016). The PG affects $\rho_c$(T) in the following way: (i) Below $T^*$, $\rho_c$(T) exhibits a sharp, pronounced upward inflection or a steep rise. Instead of continuing a smooth semiconducting growth or metallic behavior, the suppression of spectral weight across the Fermi surface (and especially around the anti-nodes in the Brillouin zone, which dominates interlayer tunneling matrix elements) severely inhibits c-axis charge transport and (ii) the resistivity anisotropy ratio, $\rho_c$(T)/$\rho_{ab}$(T), increases drastically by orders of magnitude in the pseudogap regime, reaching values from $10^3$ to well over $10^5$. This demonstrates that the PG affects interlayer charge transport far more strongly than the intralayer transport.

The temperature dependence of the Hall coefficient, $R_H$(T), is another bulk transport coefficient significantly affected by the PG (Nagaosa et al., 2010; Wang et al., 2006). The PG affects the Hall effect primarily by changing the low-energy Fermi-surface structure and its momentum-dependent spectral weight distribution and scattering, rather than simply by reducing the total number of holes (Clayhold et al., 1988; Nagaosa et al., 2010; Wang et al., 2006). The striking negative Hall coefficient observed in underdoped cuprates is strong evidence for a small electron-like contribution, generally associated with Fermi-surface reconstruction, although the precise relationship between this reconstruction and the fundamental PG remains an active research problem till date (Fang et al., 2022; Johnson et al., 2013). In the "normal" (at temperatures above $T^*$) state, to a certain extent, hole doped cuprates behave like a conventional metal with a large, hole-like Fermi surface. For hole-doped materials, the Hall number $n_H$ ($n_H = 1/eR_H$; e is the electronic charge) is expected to be roughly 1+p (where p is the doped hole content in the $CuO_2$ plane). However, as the material enters the PG regime below $T^*$, the Fermi surface reconstructs into smaller hole pockets (Fermi arcs) and sometimes electron pockets (Fang et al., 2022). When temperature is lowered below $T^*$, the Hall coefficient typically stops following its high-temperature behavior. Instead of remaining constant, $R_H(T)$ increases sharply (showing an upward turn) and, in many materials [such as $La_{2-x}Sr_xCuO_4$ (La214) and Y123)], can even change sign from positive to negative at low temperatures before the onset of superconductivity. This reconstruction is driven by the emergence of an ordering wavevector, generally associated with antiferromagnetic (AF) correlations or charge density waves (CDW) (Fradkin et al., 2015; Kivelson et al., 2003). The emergent, smaller Fermi surface volume leads to a dramatic reduction in the Hall number. Theoretically, the Hall number is found to drop from the large value down to a value roughly equal to p as the PG grows.

In the underdoped regime of high-$T_c$ cuprate superconductors, the emergence of the PG phase characterized by a partial suppression of the electronic density of states around the Fermi level above the superconducting critical temperature leaves a distinct mark on the spin magnetic susceptibility $\chi(T)$ (Naqib et al., 2009; Naqib and Islam, 2008; Islam and Naqib, 2010). Rather than exhibiting the weak, temperature-independent Pauli paramagnetic characteristic of conventional metals, the uniform spin susceptibility of cuprates starts to drop continuously well above $T_c$, starting at a characteristic temperature T* where the pseudogap makes its presence felt (Naqib et al., 2009; Naqib and Islam, 2008; Islam and Naqib, 2010).

This list of experimental probes and properties covered in this section is not exhaustive. There are several other experimental probes and properties that are affected by the PG (Tallon and Loram, 2001; Timusk and Statt, 1999; Loram et al., 1998a; Lee, 2008).

**3. Charge transport and magnetic study of the pseudogap**

In this section we focus on two of the most prominent physical properties of copper oxide superconductors – charge transport and magnetic state. The results presented herein comprises data collected from our previous studies (Naqib et al., 2005a; Naqib et al., 2005b; Naqib et al., 2003).

**3.1. In-plane and out-of-plane resistivity**

The temperature dependence of electrical resistivity, ρ(T), is one of the most useful bulk probe to understand how electrons interact with bosonic degrees of freedom, impurities, defects, and other electrons in a solid (Ashcroft and Mermin, 1976). It is particularly important for distinguishing ordinary metals from semiconductors, Fermi liquids, and quantum-critical materials. The central idea is that electrical resistivity depends on how frequently charge carriers are scattered and how effectively each scattering event relaxes electrical current. The details depend strongly on the electronic band structure and the scattering mechanisms involved (Naqib et al., 2005a; Naqib et al., 2005b; Naqib et al., 2003). The analysis is particularly challenging for strongly correlated electronic systems like hole doped cuprates where the nature of quasiparticle excitations and the nature of intertwined orders are a matter of intense debate (Naqib et al., 2005a; Naqib et al., 2005b; Naqib et al., 2003; Naqib and Islam, 2015).

PG influences the temperature dependent resistivity systematically via hole content dependent band evolution and charge carrier evolution, both of which affect the temperature dependent scattering dynamics.

In this section we present results of resistivity measurements of pure single crystals of Y123, polycrystalline Y123 and Ca doped Y123 together with those of c-axis oriented high-quality epitaxial thin films. Divalent Ca substitution in place of trivalent Y enables one to control the hole content within the $CuO_2$ planes irrespective of the oxygen deficiency (δ). In pure Y123 the hole content is controlled only via the oxygen deficiency, δ. The maximum superconducting $T_c$ in Y123 is about 93 K obtained at an optimum hole content p = 0.16 (Presland et al., 1991; Tallon et al., 1995). $T_c$ varies parabolically with p (Presland et al., 1991; Tallon et al., 1995). Superconducting transition temperature as a function of hole concentration follows a dome shaped curve starting at p ~ 0.05 in the underdoped region and finishing around p ~ 0.27 in the overdoped side. Details regarding samples, their characterization, annealing protocols, and resistivity measurement can be found elsewhere (Naqib et al., 2005a; Naqib et al., 2005b; Naqib et al., 2003; Naqib and Islam, 2015). The hole content of all the compounds were determined using two different measurements – room temperature thermopower (Obertelli et al., 1992) and the widely applied $T_c$ versus p scaling (Presland et al., 1991; Tallon et al., 1995).

Figs. 1 show the systematic changes in ρ(T) with p. ρ(T) for samples with p > 0.19 gradually develops an increasingly positive curvature as T decreases whereas a negative curvature grows with decreasing p for the underdoped (UD) compounds. The slope of ρ(T) increases systematically as p is reduced. The residual resistivity also increases monotonically with underdoping. The evolution of ρ(T) with p provides a way of establishing the T-p phase diagram of high-$T_c$ superconductors and can give measures of $T^*(p)$, the characteristic pseudogap temperature. Conventionally $T^*$ is defined as the temperature at which ρ(T) starts to decrease at a faster rate than from its high-T linear behavior (Naqib et al., 2005a; Naqib et al., 2005b; Naqib et al., 2003; Naqib and Islam, 2015). One particular problem associated with this method is that near the vicinity of the optimum doping level and above, $T^*(p)$ is quite close to $T_c(p)$ and the fall of ρ(T) due to strong superconducting (SC) fluctuation masks any possible gradual reduction in ρ(T) due to the presence of the PG as observed at higher-T in the UD compounds. The proposed scenarios to explain the origin and the p-dependence of the pseudogap could be classified mainly into two broad categories (Tallon and Loram, 2001; Timusk and Statt, 1999). The first one is based on the precursor pairing scenario, where the pseudogap is derived from strong fluctuations of SC origin for systems with low dimensionality and low superfluid density. In the second scenario, the pseudogap is attributed to correlations of some other (non-superconducting) type, which coexists [i.e., $T^*(p)$ exists below the $T_c(p)$ line] and often competes with superconductivity. The experimental situation is often thought to be rather inconclusive regarding the origin of the pseudogap (Tallon and Loram, 2001; Timusk and Statt, 1999). The close proximity between $T^*$ and $T_c$ near and above the optimum hole concentration does not pose any problem for theories belonging to the first group where $T^*(p)$ itself is essentially derived from strong superconducting fluctuations and $T^*(p)$ merges with the $T_c(p)$ in the overdoped region (OD) region ($p \geq 0.21$) where the pair formation temperature ($T^*$) is essentially the same as the phase coherence temperature (the temperature at which there is long-range order and the superconducting transition takes place in the thermodynamical sense, i.e., $T_c$). For the second scenario, the closeness of $T^*$ and $T_c$ poses a serious problem as superconducting pairing fluctuations (and superconductivity itself) mask the signatures of the predicted pseudogap in the vicinity of (and below) $T_c$.

In this study, two methods have been employed to determine $T^*$. As can be seen from Fig. 2a, representative plots of $d\rho(T)/dT$ versus T and $[\rho(T) - \rho_{LF}]$ versus T yield very similar $T^*$ values (within ± 5K). Here $\rho_{LF}$ is a linear fit of the form $\rho_{LF} = b + cT$, in the high-temperature region of the $\rho(T)$.

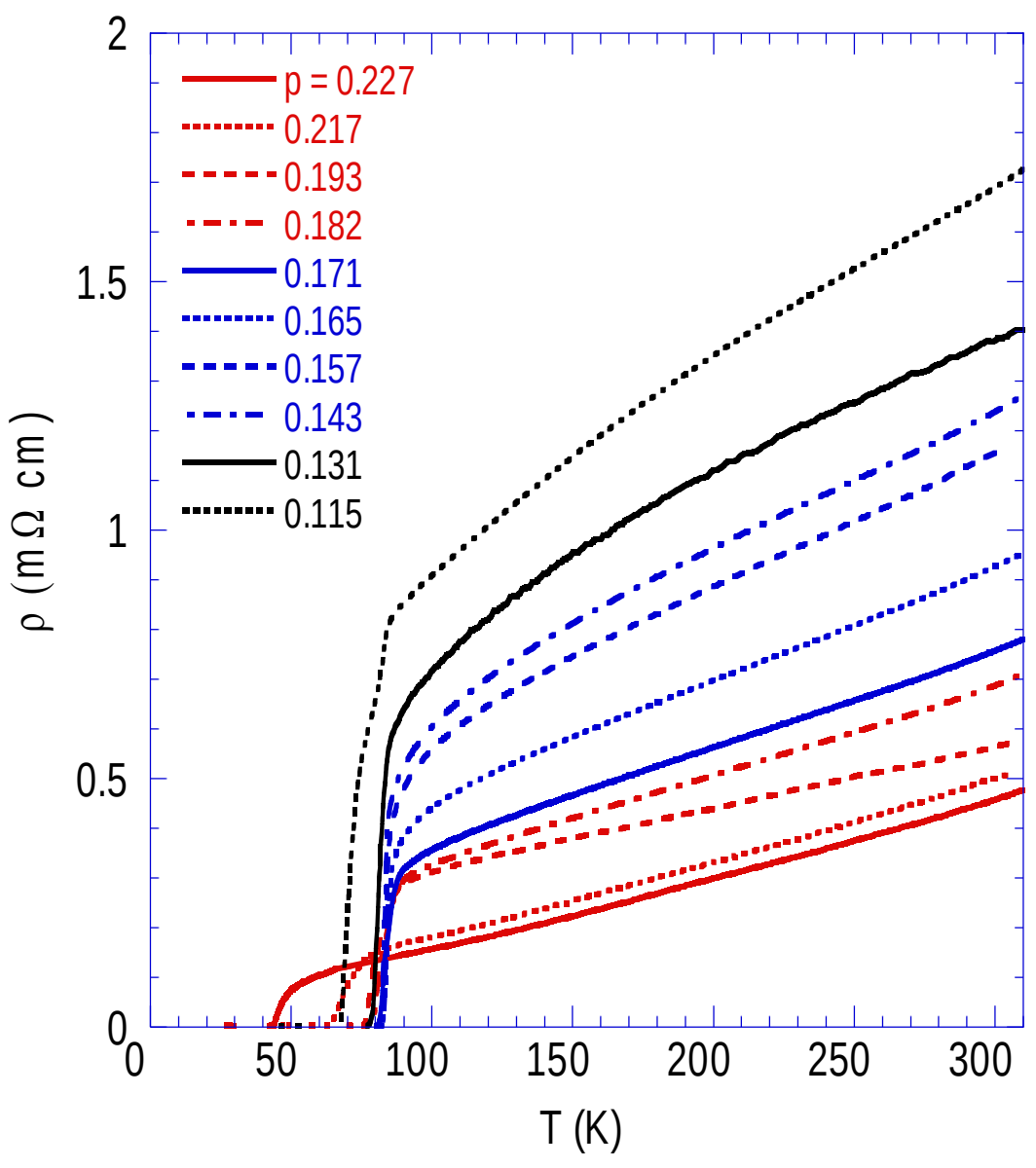

Fig. 1a: Resistivity of sintered $Y_{0.80}Ca_{0.20}Ba_2Cu_3O_{7-\delta}$. Hole contents (within ± 0.004) are shown in the plot.

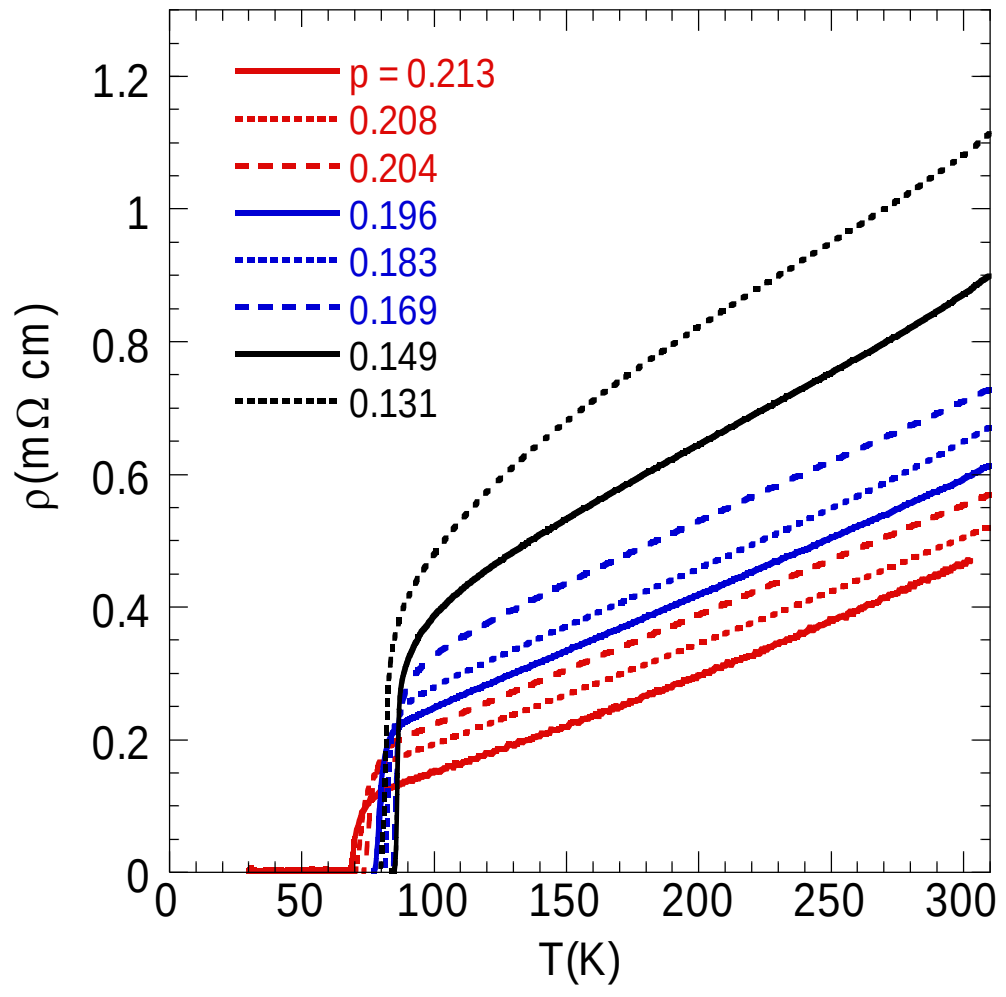

Fig. 1b: Resistivity of sintered $Y_{0.90}Ca_{0.10}Ba_2Cu_3O_{7-\delta}$. Hole contents (within ± 0.004) are shown in the plot.

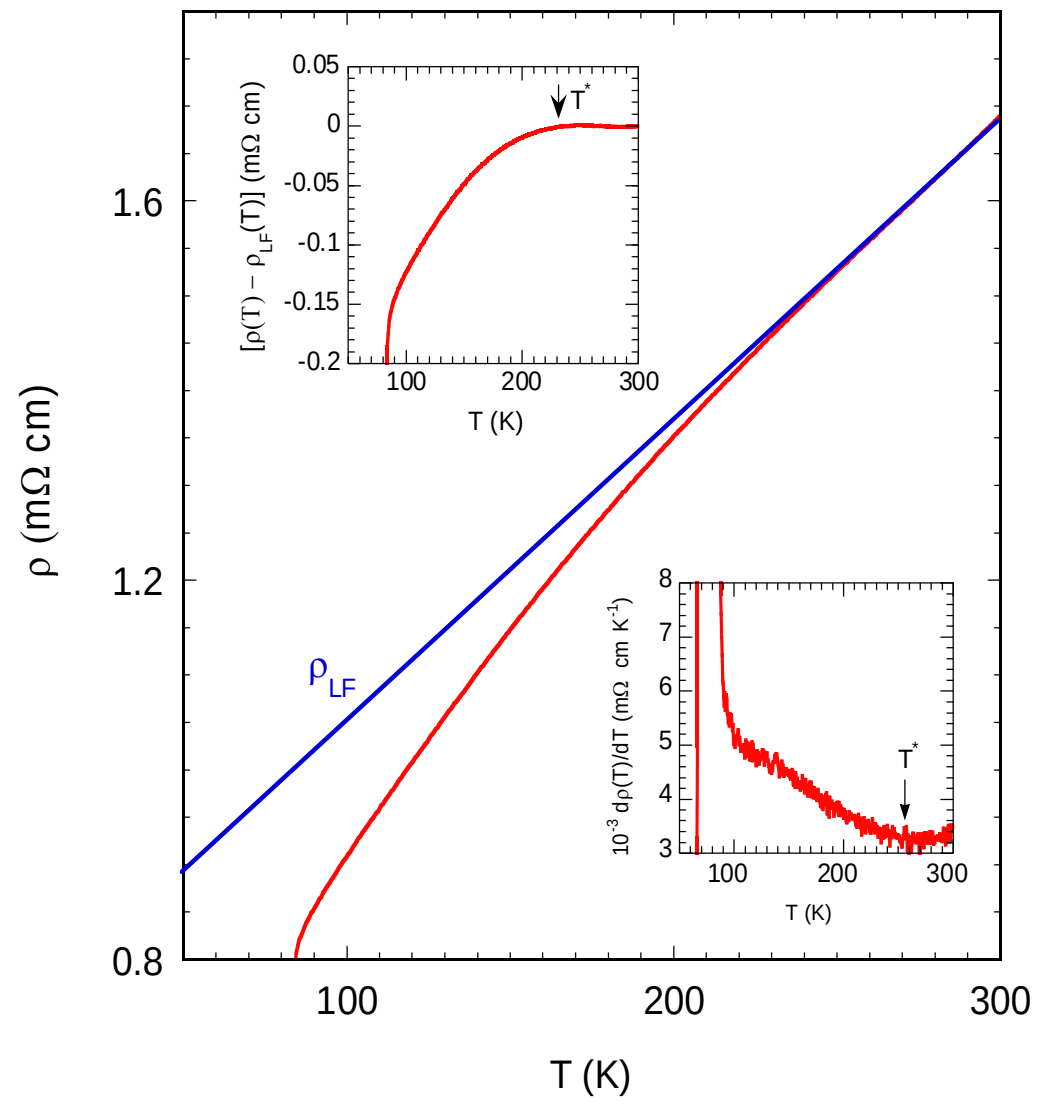

Fig. 2a: Methods employed to determine $T^*$ for $Y_{0.80}Ca_{0.20}Ba_2Cu_3O_{7-\delta}$ (p = 0.115) (see text for details).

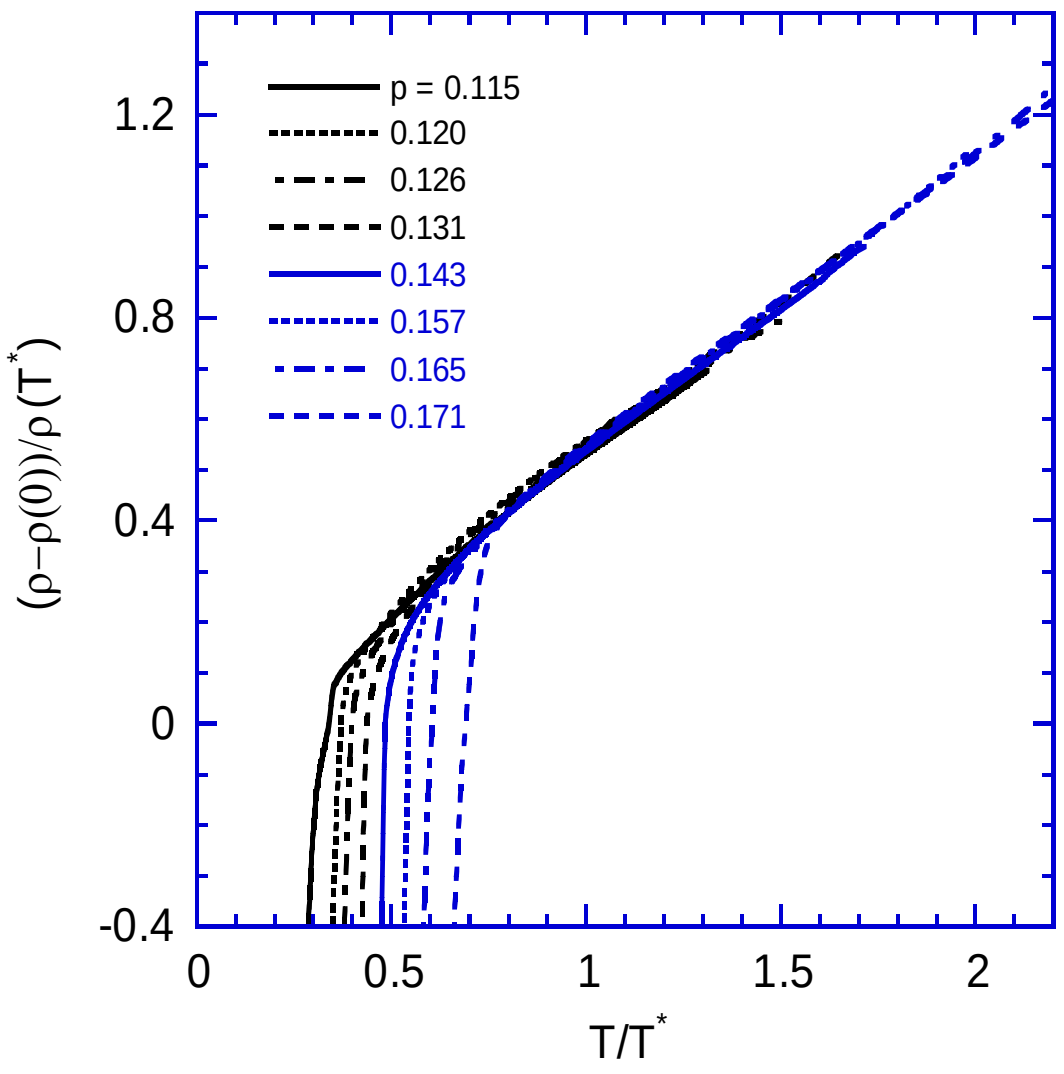

Fig. 2b: Normalized resistivity, with $T^*(p)$ as scaling parameter for $Y_{0.80}Ca_{0.20}Ba_2Cu_3O_{7-\delta}$. p-values (within ± 0.004) are shown in the plot. ρ(0) is the residual resistivity.

Also using ($T^*$/T) as a scaling parameter, it is possible to present ρ(T) for different hole contents. The result of such scaling is shown in Fig. 2b, where, leaving aside the superconducting transitions, all resistivity curves collapse onto one p-independent universal curve over a wide temperature range, signifying the same dominant electronic scattering dynamics characterized by the presence and extent of the PG energy.

This scaling also shows that the extent of the PG ($T^*$) sets the dominant energy scale for low-lying QP excitations in Ca substituted Y123 for compounds with $p < 0.18$. It is worth mentioning that the departure from the scaling invariably starts at a temperature ~ 20 ± 5K above $T_c$. At this temperature strong SC fluctuations are expected due to short-lived Cooper pairs (Naqib et al., 2005a; Naqib et al., 2005b; Naqib et al., 2007). The validity of the scaling at higher temperature and not below ~ 20K above $T_c$ points towards different origin for the gradual downturn in the resistivity at $T^*$ and that in the vicinity of $T_c$. $T^*(p)$ obtained from such scaling agrees quite well with those obtained by using the other two methods (shown in Fig. 2a). It is interesting to note that, since ρ(T) of samples with p > 0.19 develops increasingly positive curvatures with increasing overdoping (Figs. 1), a scaling with any kind of characteristic temperature will deviate from the universal scaling shown in Fig. 2b. This could be an indication of a fundamental change in the electronic ground states of cuprates in the OD side (p > 0.19), namely, the absence of a pseudogap in the electronic density of states.

Fig. 3 shows some representative in-plane resistivity [$\rho_{ab}$(T)] data for high-quality epitaxial $Y_{1-x}Ca_xBa_2Cu_3O_{7-\delta}$ thin films with different values of hole and Ca contents (Naqib et al., 2005a; Naqib et al., 2005b; Naqib et al., 2004). It is found that, for a given value of p, $T^*$ is independent of the crystalline state of the samples – be it polycrystalline, epitaxial thin films or single crystals (Naqib et al., 2005b; Naqib et al., 2004). This is shown in Fig. 4, which also demonstrates clearly that for sintered samples, transport charge is mainly governed by the ab-plane charge scattering mechanisms. This follows from the fact that there is a large resistivity anisotropy in quasi two-dimensional cuprates. In-plane charge transport dominated by the conducting $CuO_2$ layers is much stronger than that of the out-of-plane one. Thus, the c-axis resistivity ($\rho_c$) is significantly higher than $\rho_{ab}$. This is a consequence of the generic structural feature of all cuprate superconductors. The resistivity anisotropy increases rapidly as hole content decreases. In other words, the $CuO_2$ planes become more and more decoupled with underdoping (Naqib et al., 2011). The schematic structure of Y123 is displayed in Fig. 5. At this point it is worth noting that the systematic evolution (super-linear → linear → sub-linear) of the ρ(T, p) with decreasing hole concentration has an electronic origin since the phonon spectrum does not vary significantly with p in cuprate superconductors. This may also suggest that phonons are not strongly influencing the charge scattering processes and therefore, superconductivity in cuprates originates mainly from non-phonon mechanisms.

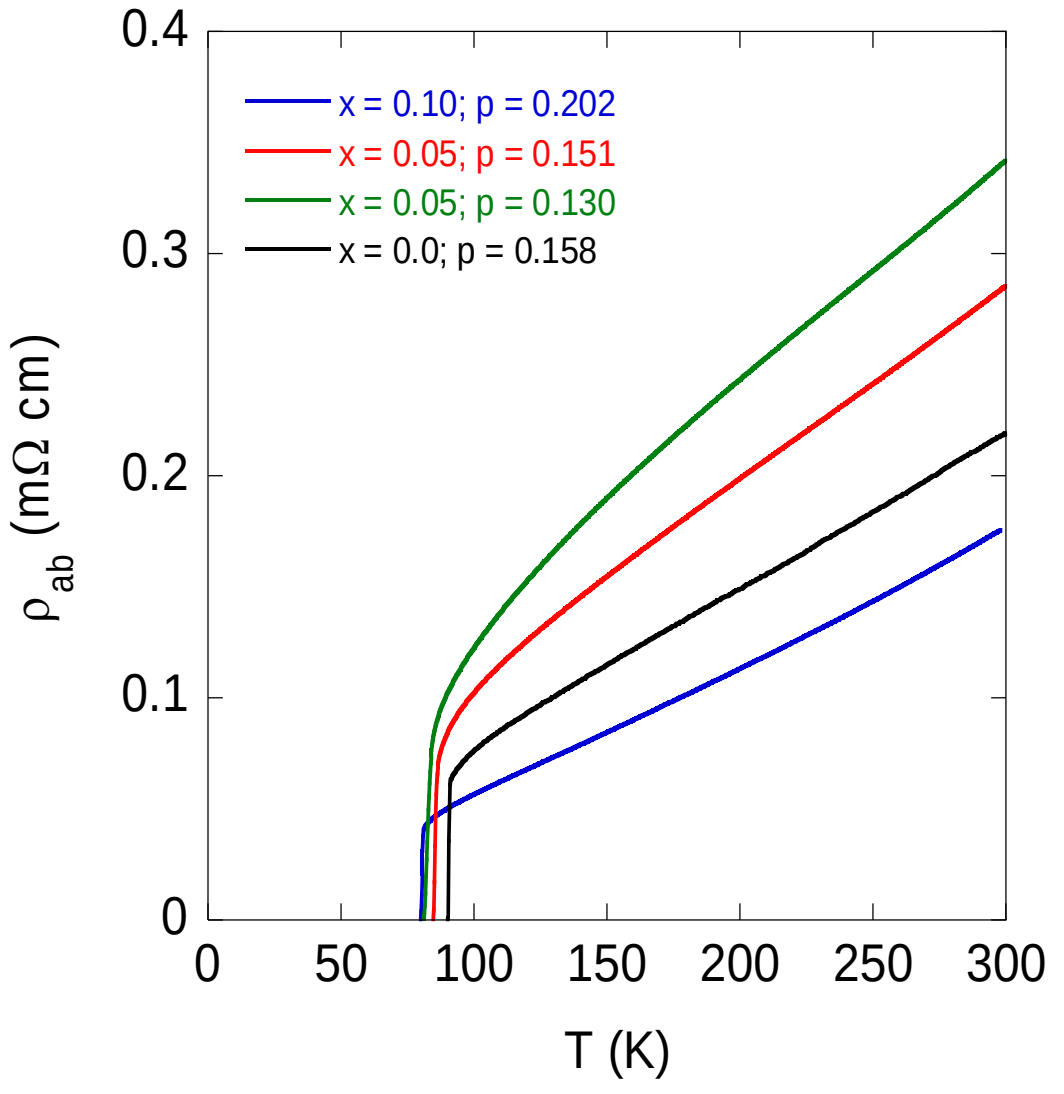

Fig. 3: In-plane resistivity of c-axis oriented $Y_{1-x}Ca_xBa_2Cu_3O_{7-\delta}$ thin films. Ca contents and p-values (within ± 0.004) are given.

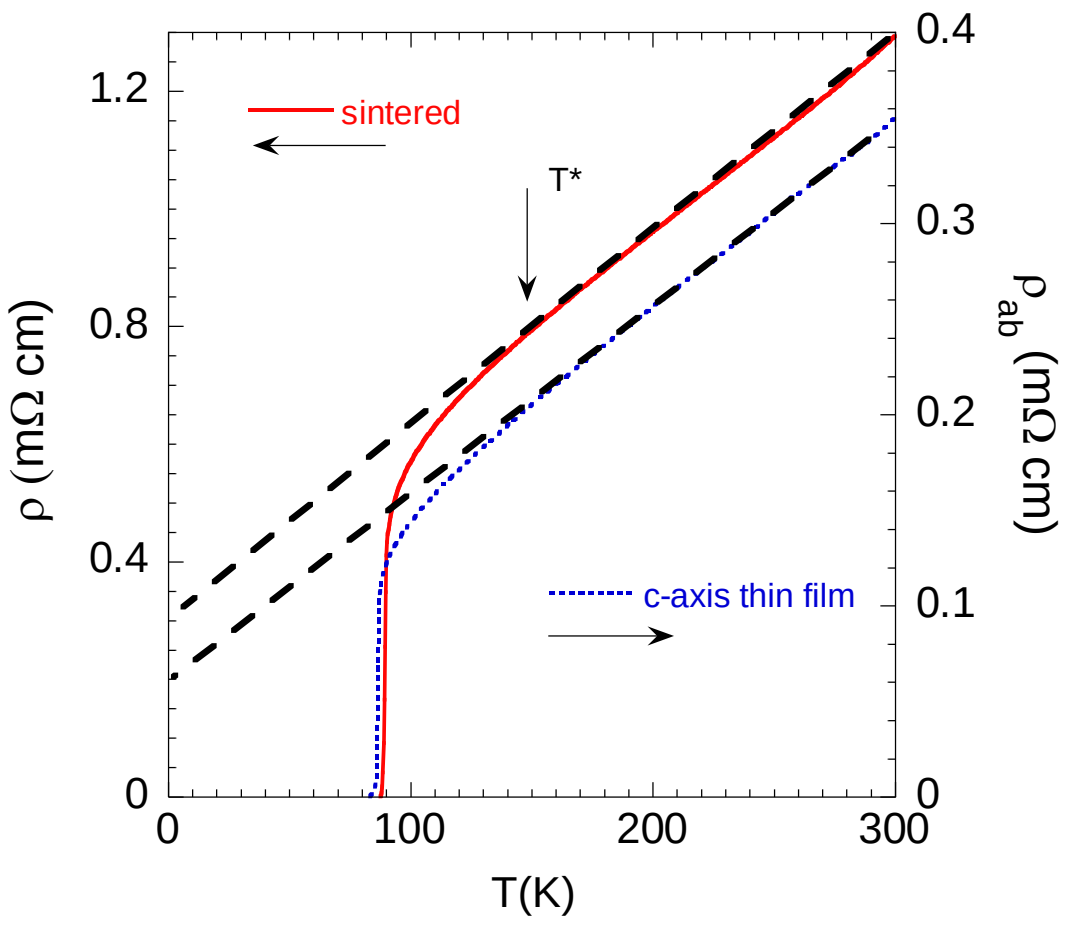

Fig. 4: Resistivity and T* of sintered and c-axis oriented $Y_{0.95}Ca_{0.05}Ba_2Cu_3O_{7-\delta}$ compounds for p = 0.155 ± 0.003.

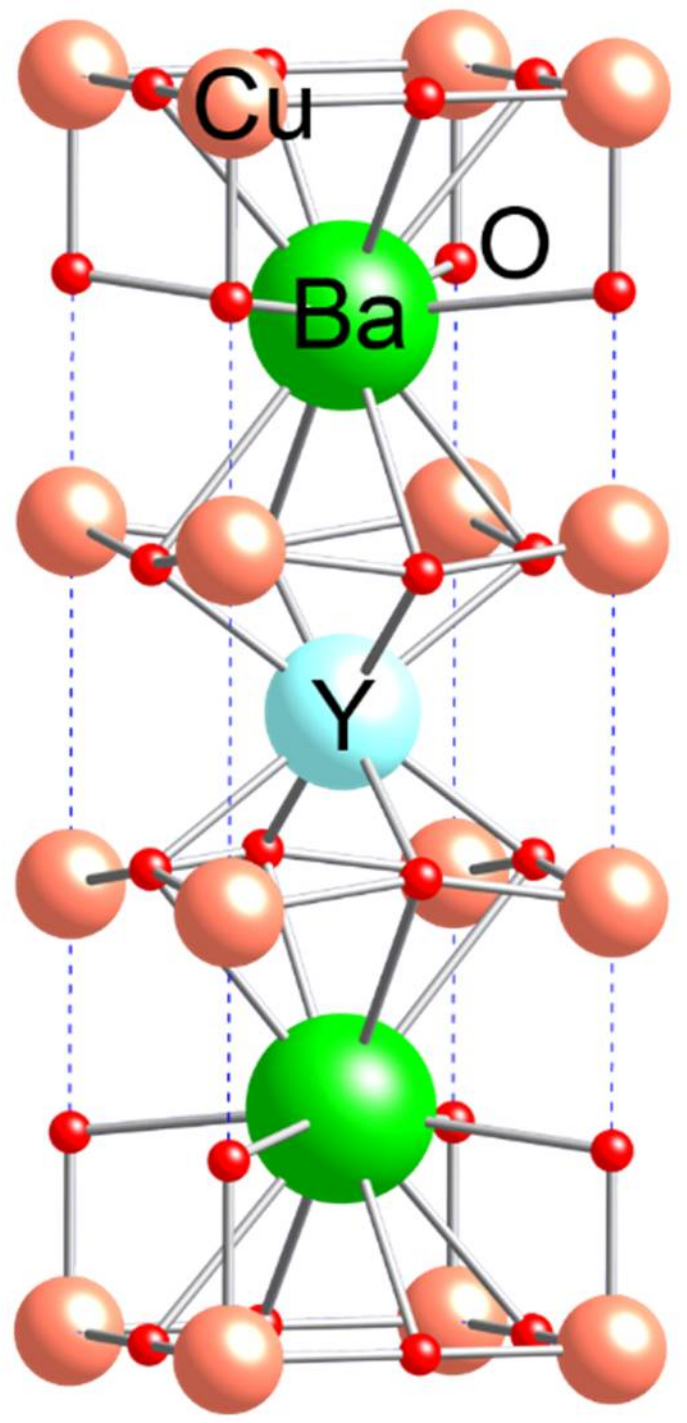

Fig. 5: Schematic crystal structure of Y123. The conducting Cu-O planes are separated by the Ba-O layers serving as charge reservoirs.

To illustrate the completely contrasting nature of the charge transport along c-axis, we have shown the hole content dependent $\rho_c(T)$ in Fig. 6 (Naqib et al., 2011).

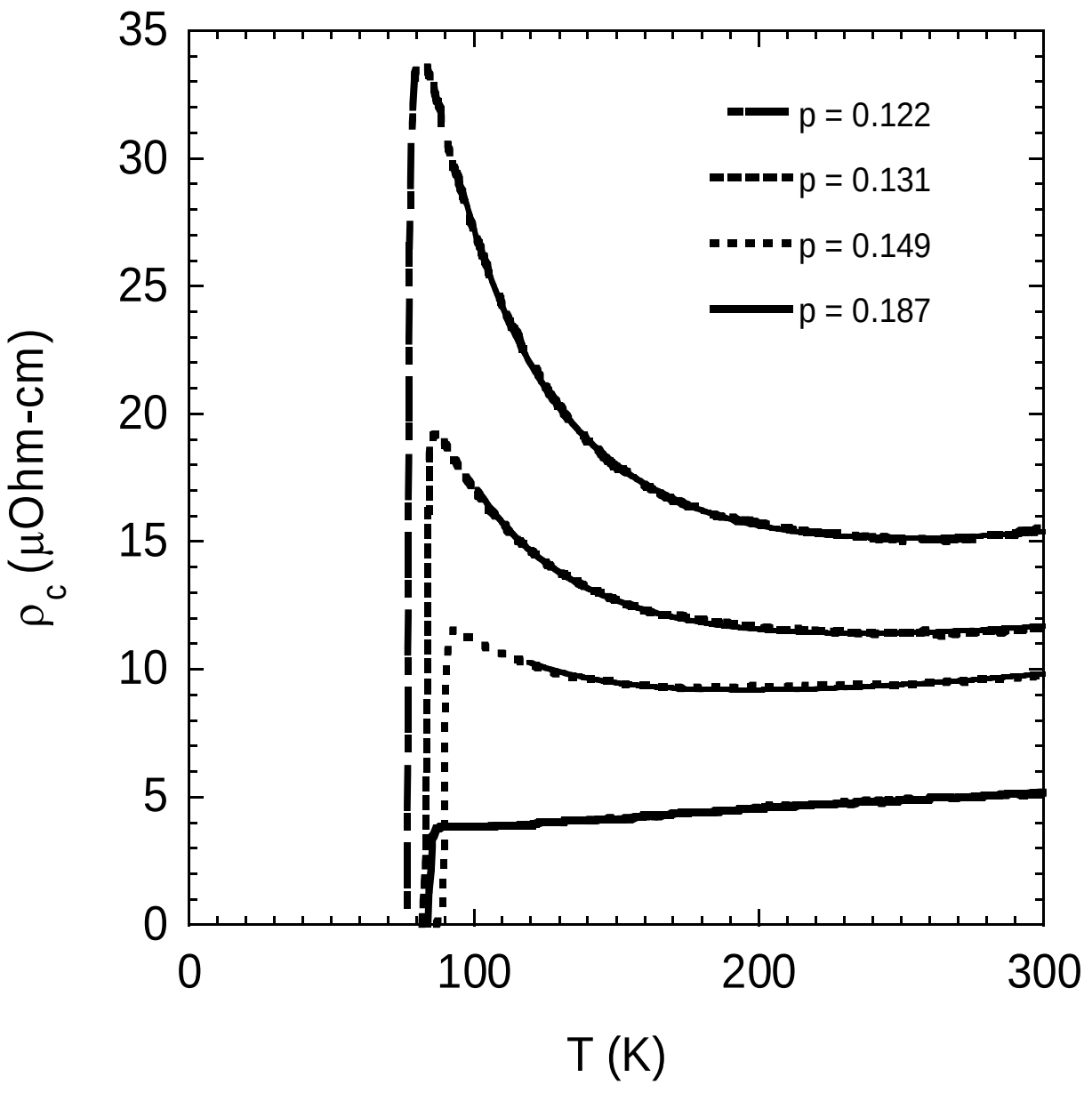

Fig. 6: $\rho_c$(T) of $YBa_2Cu_3O_{7-\delta}$ single crystals. Hole contents (within ± 0.004) are shown in the plot.

The PG induces a downturn in $\rho_{ab}$(T). The effect of PG is opposite in $\rho_c$(T). PG induces an upturn in $\rho_c$(T). This upturn sets at higher temperatures and becomes stronger with the depletion of hole concentration within the $CuO_2$ planes. The c-axis resistivity upturn in UD cuprates is a direct consequence of interlayer incoherence coupled with pseudogap-induced spectral weight depletion at the anti-nodes in the Brillouin zone. Because c-axis transport relies on charge carriers tunneling between weakly coupled $CuO_2$ planes, the opening of the pseudogap removes the low-energy electronic states particularly from the anti-nodal regions of the momentum space which are required for this tunneling. This causes out-of-plane charge transport to exhibit insulating-like growth while in-plane transport remains metallic. Close to p ~ 0.19, both in-plane and out-of-plane resistivity becomes completely linear in temperature except at temperature where strong SC fluctuations set in (Naqib et al., 2005a; Naqib et al., 2005b; Naqib et al., 2003; Naqib and Islam, 2015). This strange T-linear resistivity suggests non-Fermi liquid behavior and possibly has profound implication on the anomalous normal and superconducting state properties of high-$T_c$ cuprates (Loram et al., 2001; Tallon et al., 1999; Panagopoulos et al., 2002; Naqib et al., 2003). We will discuss this particular aspect in Section 4.

### 3.2. Magnetic susceptibility

One of the most important information that can be extracted from the analysis of temperature and hole concentration dependent magnetic susceptibility [$\chi$(T, p)] is about the temperature and p dependences of low-energy electronic density of states of hole doped copper oxide superconductors. As EDOS, as a function of energy (or, equivalently, temperature), is at the center of any problem associated with PG, $\chi$(T, p) is indeed a powerful experimental probe to study various p-dependent features of this phenomenon. It is in fact the intrinsic spin part (the Pauli spin susceptibility, $\chi_{spin}$) of $\chi(T, p)$ that represents the QP spectral density near the Fermi-level. The Pauli spin susceptibility arises from the coupling of intrinsic spins of the mobile carriers with the applied magnetic field and, for ordinary Fermi-liquids, can be expressed as:

$$\chi_{spin}(T) = \mu_B^2 \langle N(\varepsilon_F) \rangle_T \qquad (1)$$

where, $\mu_B$ is the Bohr magneton and $\langle N(\varepsilon_F) \rangle_T$ is the thermally averaged EDOS at the Fermi-energy. In cuprates, where strong electronic correlations are present, $\chi_{spin}$ at a particular temperature T still approximately represents the thermal average value of EDOS near the Fermi-level over an energy width of $\sim \varepsilon_F \pm 2k_BT$ (Naqib and Islam, 2008).

The experimental uniform magnetic susceptibility data consists of a number of contributions coming from different origins. It is therefore important that we are able to separate these different contributions from each other so that only the terms directly related to the PG (i.e., to the QP excitation spectrum) can be considered. With all these contributions to the uniform susceptibility, we can express $\chi(T)$ as follows (Naqib and Islam, 2008):

$$\chi(T) = \chi_{spin}(T) + \chi_{Curie}(T) + \chi_{core} + \chi_{vv} + \chi_{impurity}(T) \qquad (2)$$

where the Larmor or the core susceptibility and the Van Vleck (VV) susceptibility are expected to be p- and T-independent over the experimental temperature range (Walstedt et al., 1992; Allgeier and Schilling, 1993). $\chi_{Curie}(T)$ is the paramagnetic Curie susceptibility. We have ignored Landau diamagnetic susceptibility, which has the same T-dependence as $\chi_{spin}(T)$, but is only from 2% to 5% of the $\chi_{spin}(T)$ in magnitude (Walstedt et al., 1992; Allgeier and Schilling, 1993). $\chi_{imp}(T)$ is the contribution from possible impurity phases present in the compounds. Details regarding susceptibility measurements are reported elsewhere (Naqib and Islam, 2008).

Figs. 7 show the experimental $\chi(T)$ and $\chi(T)T$ data for $Y_{1-x}Ca_xBa_2Cu_3O_{7-\delta}$ compounds and readily illustrate two important points. The first one, related to the existence of the pseudogap, is the fact that both $\chi(T)$ and $\chi(T)T$ becomes strongly p-dependent only for $p < 0.19$ ($\pm$ 0.005). The second important finding is the fact that Ca modifies the normal state magnetic properties of $Y_{1-x}Ca_xBa_2Cu_3O_{7-\delta}$ compounds quite drastically. This is seen from the progressive growth of Curie-like behavior in $\chi(T)$ with increasing Ca contents. $\chi(T)$ for Ca-substituted Y123 shows features similar to Co- or Ni-substituted Y123 (Cooper and Loram, 1996). This is quite striking, because unlike Co and Ni, Ca is non-magnetic. In fact, $Ca^{2+}$ has a full shell electronic configuration ($3p^6$), and no Curie-like contribution to the magnetization is expected. This anomalous Ca-induced contribution to the magnetic susceptibility has been investigated in a previous work (Naqib et al., 2009) and it has been found that this effect is not related to the presence or magnitude of the pseudogap.

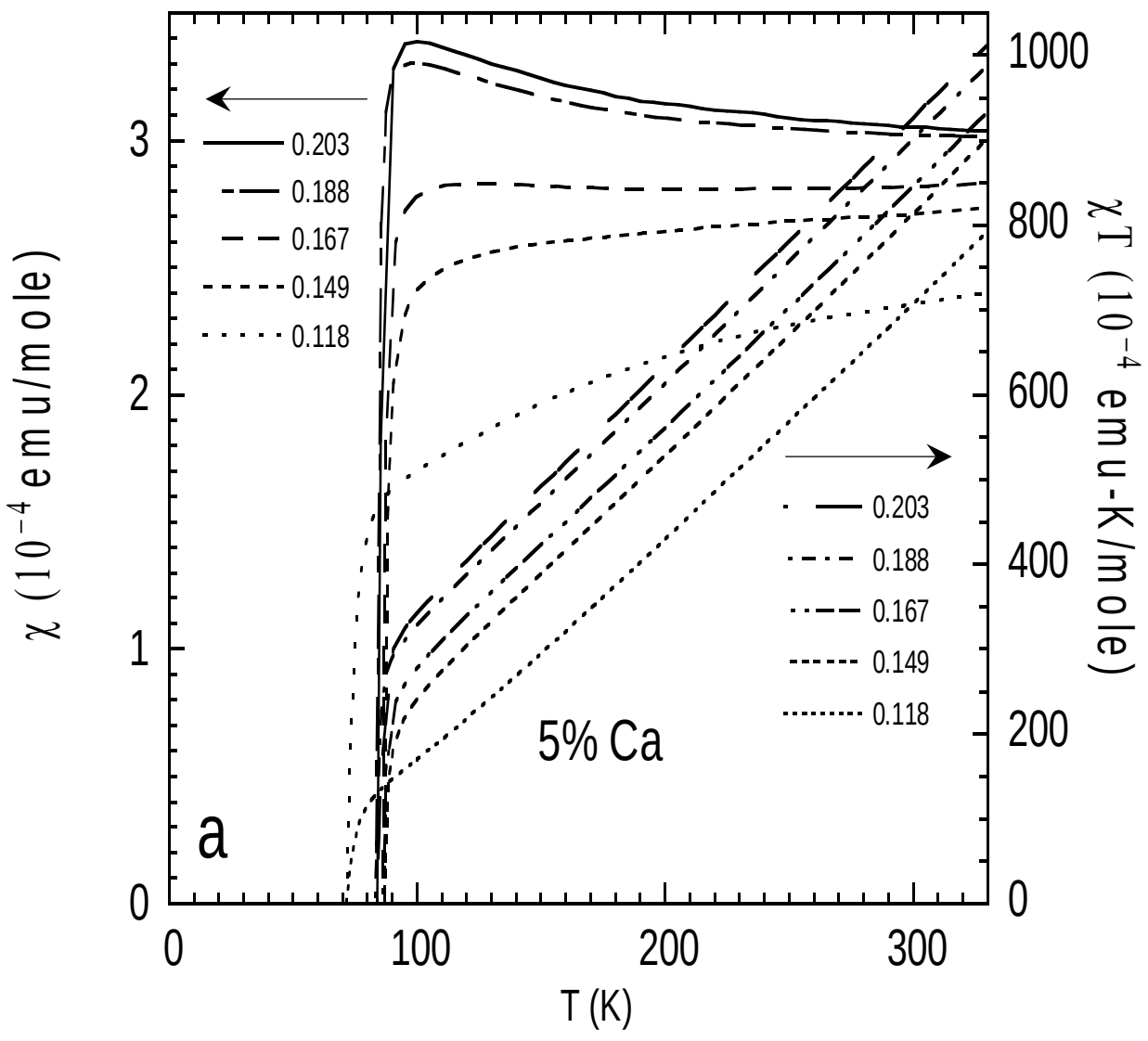

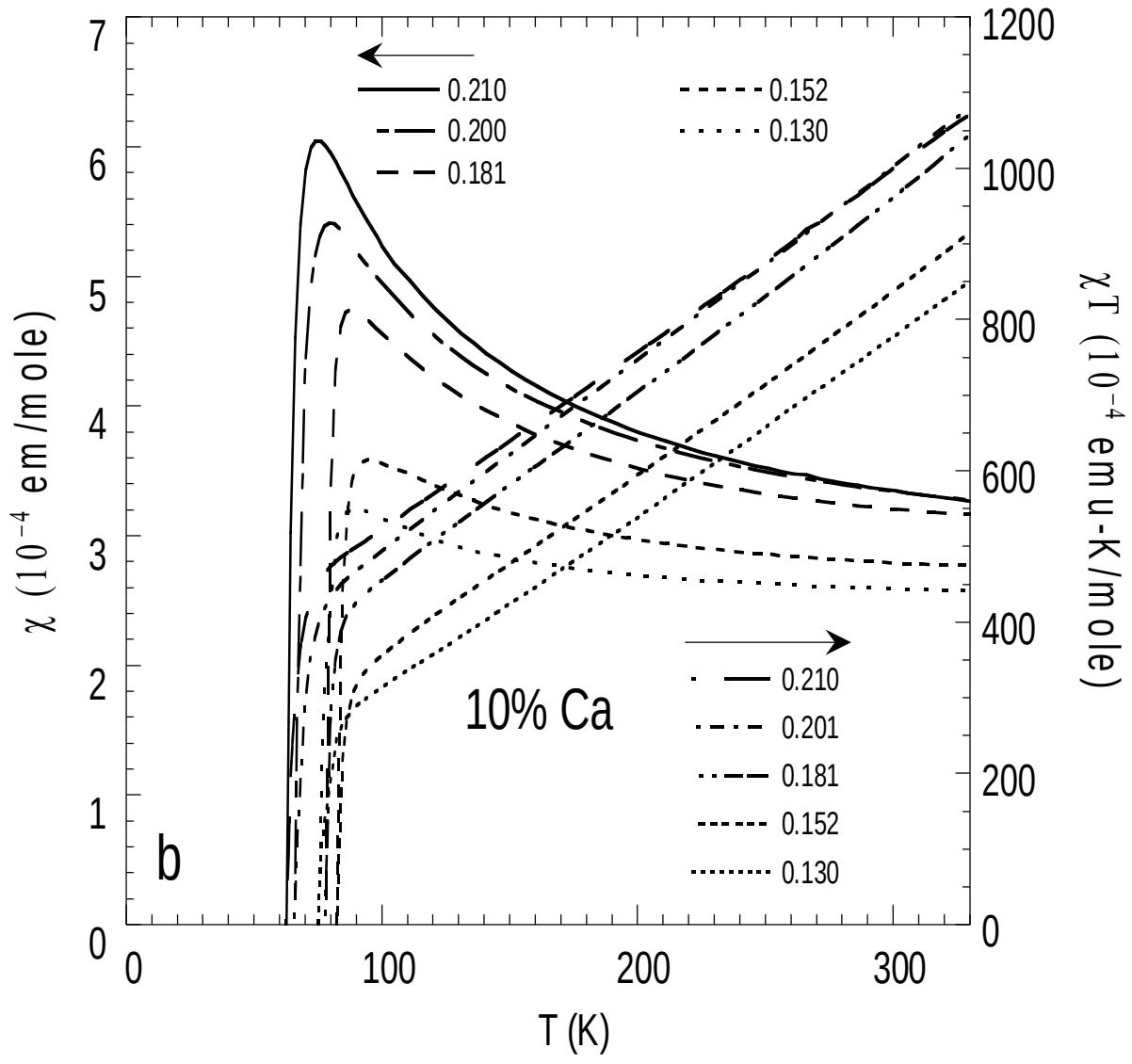

It is also seen from Figs. 7 that $\chi(T)T$ deviates from linearity at lower temperatures above $T_c$ even for samples with no pseudogap ($p > \sim 0.19$). Namely $\chi(T)T$ rises above the high-temperature linear behavior for the Ca-substituted OD compounds, implying the presence of a Curie-Weiss-like contribution to the susceptibility. This is due to the presence of small quantity of an unwanted p-independent magnetic impurity phase in these compounds, namely $BaCuO_{2+z}$ (Naqib et al., 2009).

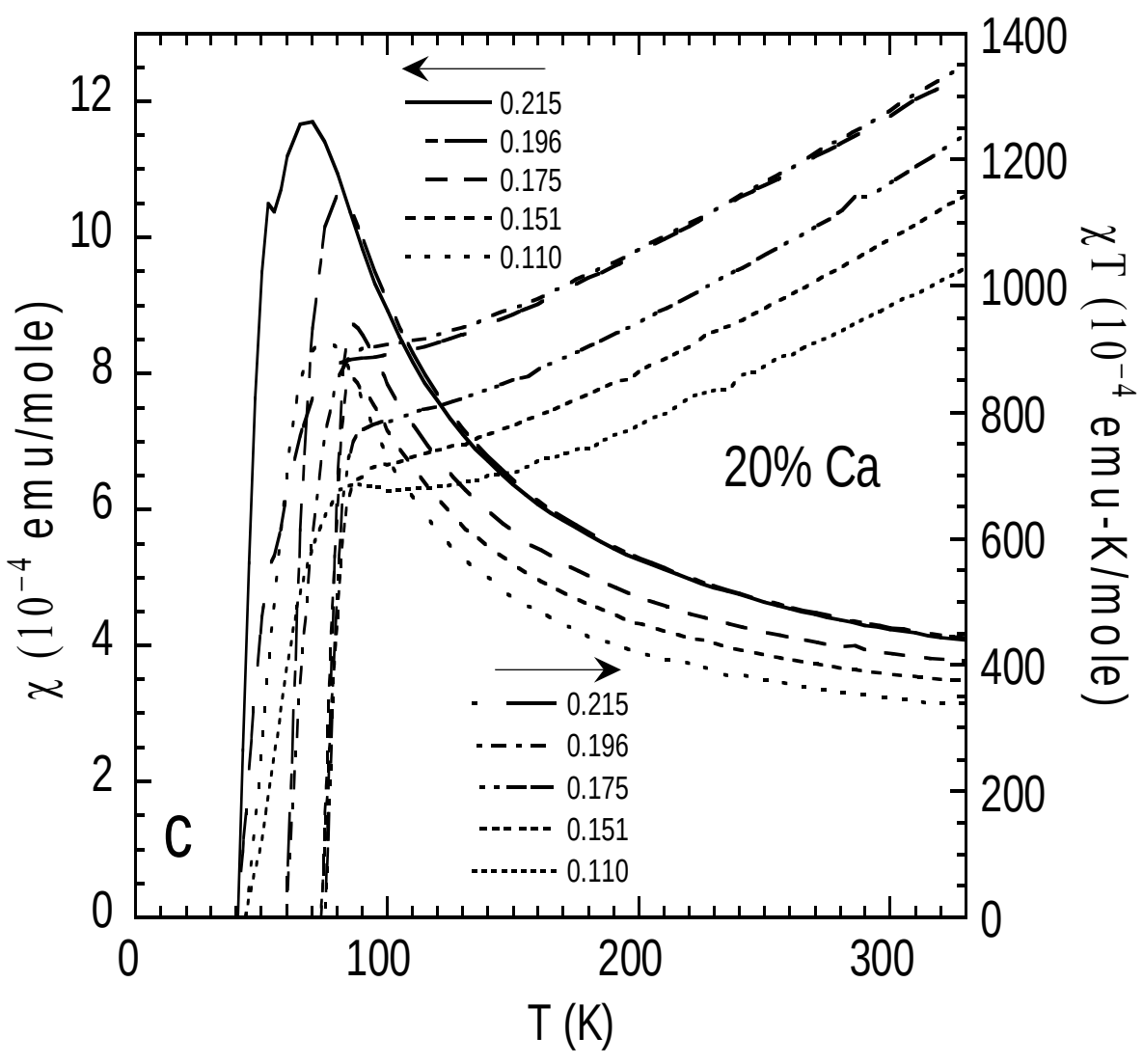

Fig. 7 (a – c): χ(T) and χ(T)T for Ca doped Y123 (p-values are accurate within ± 0.004). Ca and hole concentrations are shown.

It is important to notice that high-T linear extrapolation of χ(T)T for pure Y123 yields a negative intercept at T = 0K, as seen from Fig. 8. This is due to the presence of an approximately triangular non-states conserving PG (Naqib et al., 2009; Naqib and Islam, 2008). Such a PG introduces a negative Curie-like term in χ(T). As the width of the PG increases with underdoping, this T = 0K intercept becomes more and more negative. Positive intercept-values for Ca substituted Y123 with p < 0.19 does not imply the absence of PG for these compounds, rather, it represents a real additional (to that for pure Y123) paramagnetic Curie contribution in χ(T) due to Ca substitution. In our previous investigations, we have shown that Ca itself does not affect the magnitude of the pseudogap (Naqib et al., 2009).

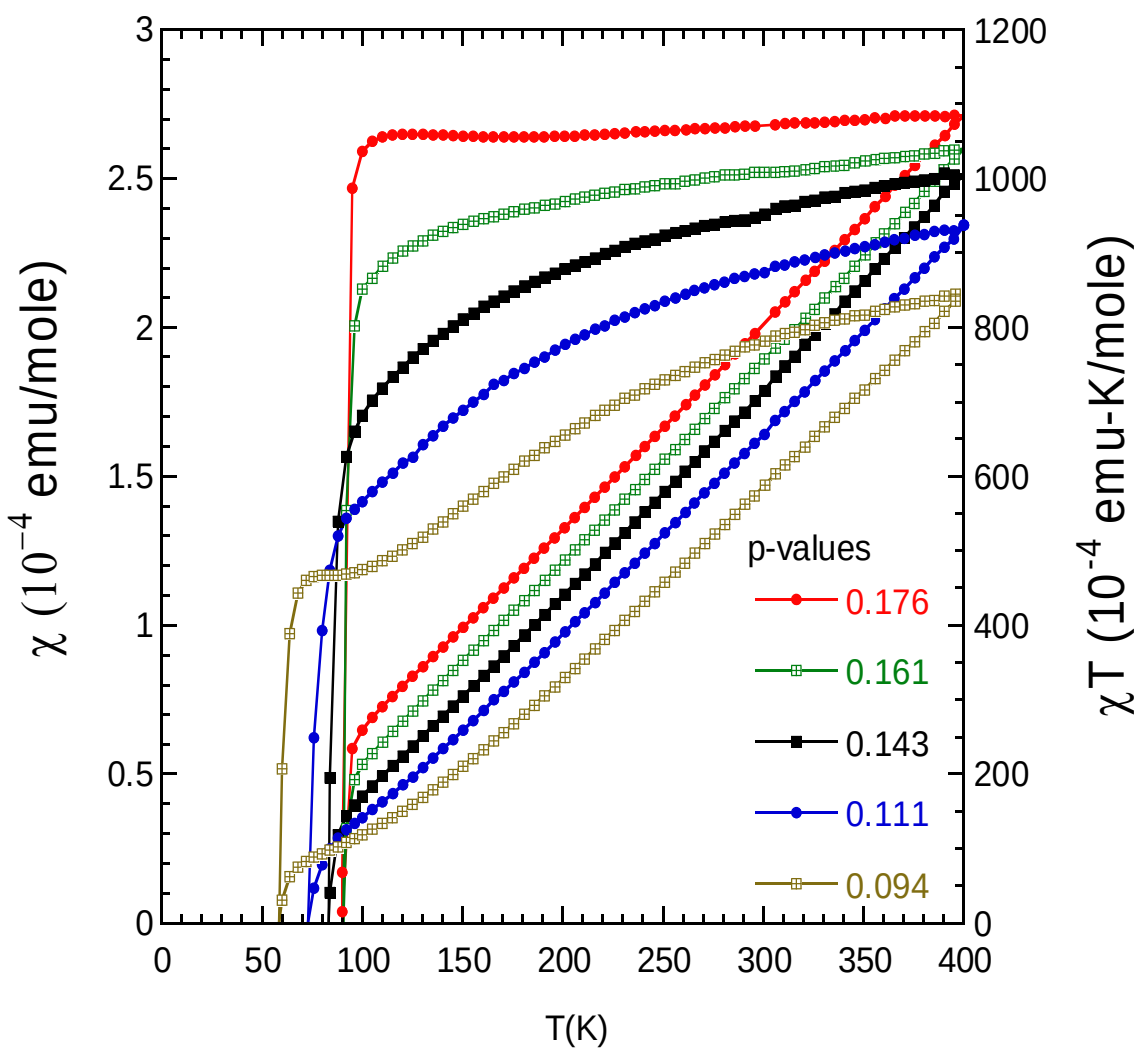

Fig. 8: χ(T) and χ(T)T for Y123 (p-values are accurate within ± 0.004). Hole concentrations are shown.

It is possible to estimate $T^*$, or equivalently, the characteristic PG energy, $E_g$, from the analysis of the uniform magnetic susceptibility data. To extract the characteristic PG energy scale, $E_g$ (= $k_BT^*$) we have used a V-shaped gap centered at the Fermi-level. In this scheme the thermally averaged electronic energy density of states can be expressed as (Naqib and Islam, 2008):

$$\langle N(\varepsilon)\rangle_T = N_0[1 - D^{-1}\ln\{\cosh(D)\}] \qquad (3)$$

where, $D = E_g/2k_BT$. Eqn. (3) captures the essential features of the PG on the electronic entropy (Cooper and Loram, 1996; Loram et al., 1998a). Fig. 9 shows results of the fits of χ(T, p) data for pure (0%Ca) Y123 samples from 400K to ~ $T_c$ + 30K (the lower temperature limit has been set to avoid significant diamagnetic SC fluctuations close to $T_c$) using Eqns. (1) and (3). An excellent fit to the experimental magnetic susceptibility data is obtained as illustrated in Fig. 9.

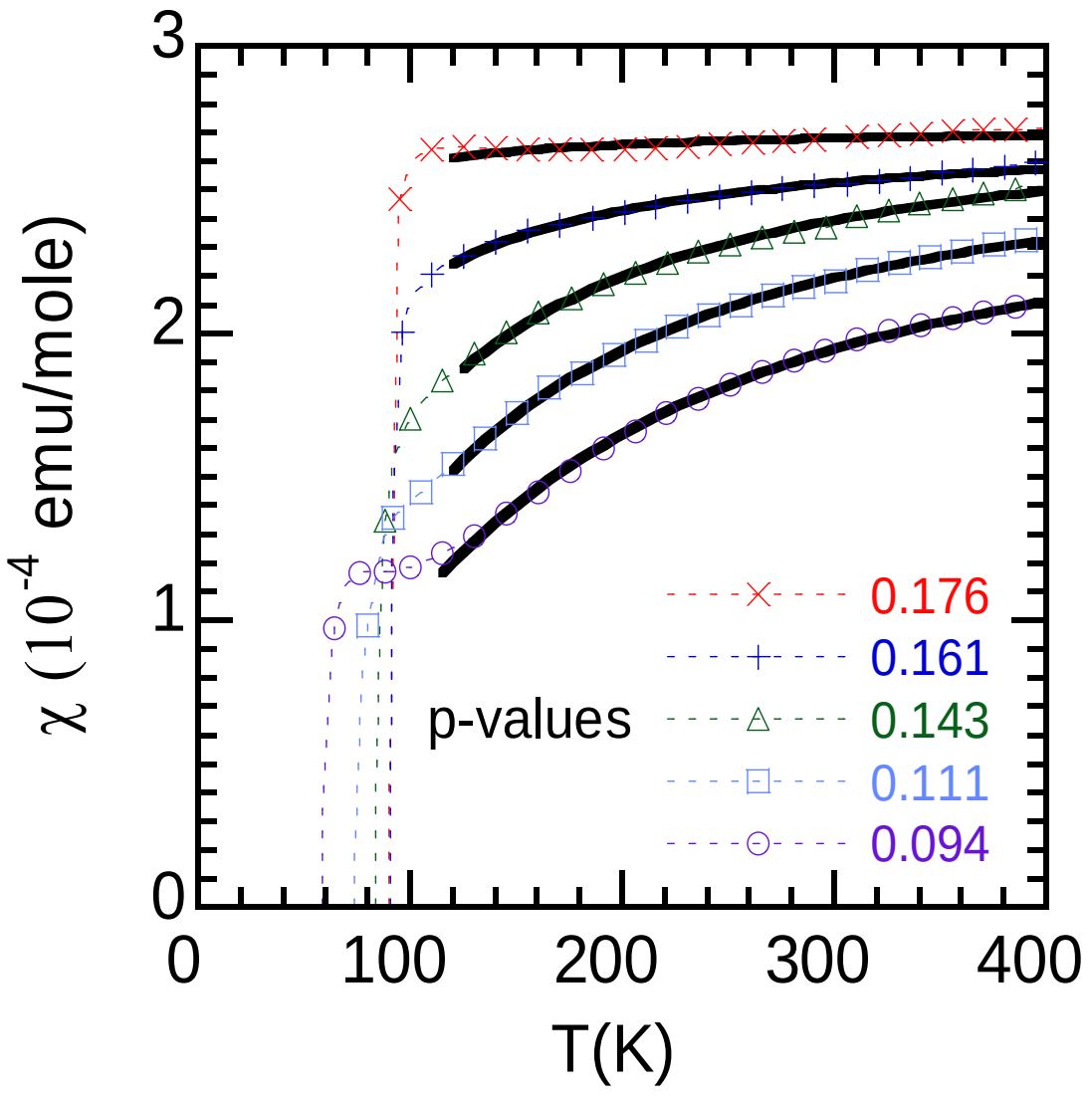

Fig. 9: χ(T, p) of pure Y123. Thick full lines show the fits to Eqn. 3.

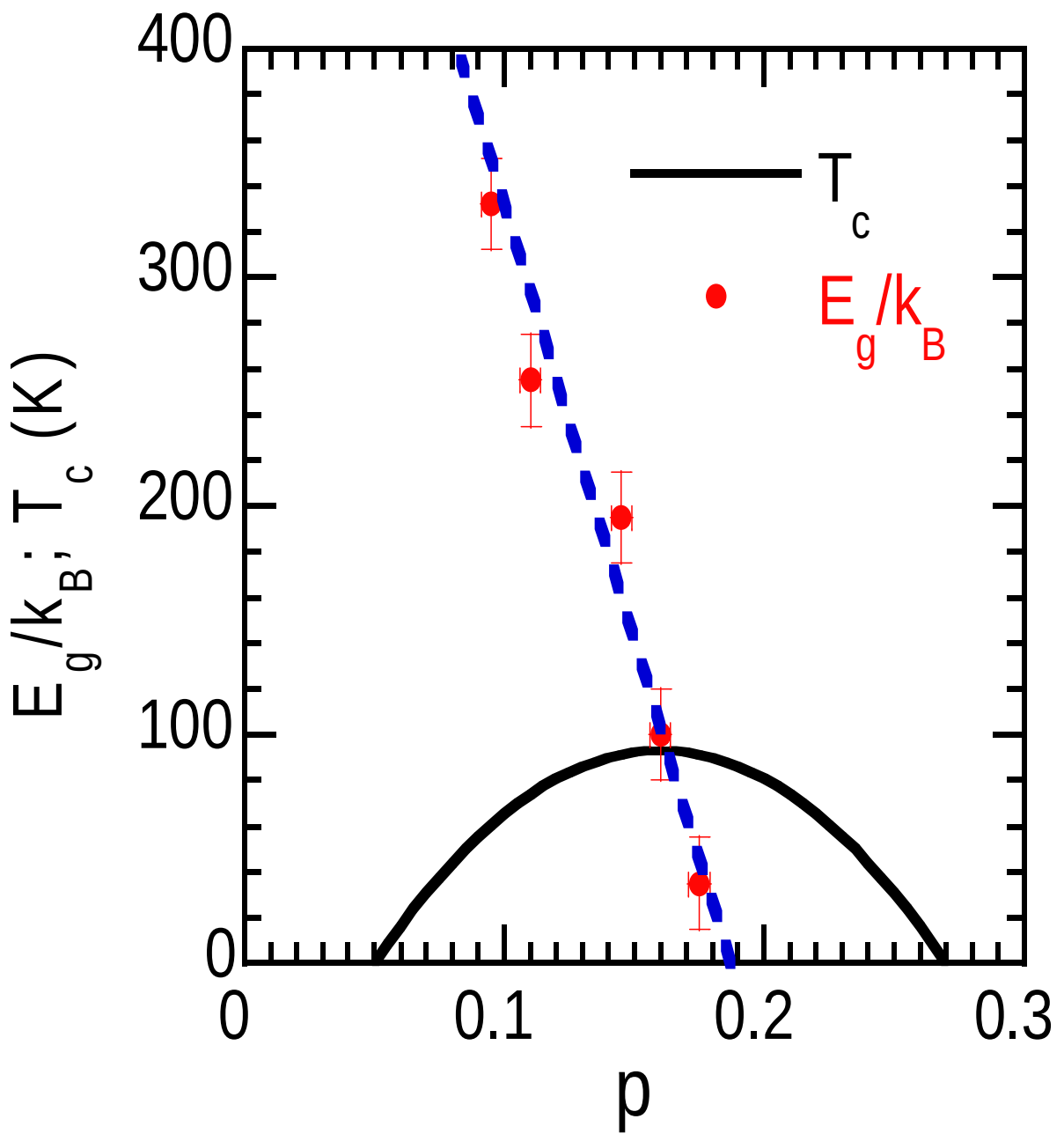

Fig. 10: $E_g(p)/k_B$ and $T_c(p)$ of pure Y123. The dashed line is drawn as a guide to the eyes. The $T_c$ – p is constructed for Y123 using the parabolic relation with a maximum $T_c$ of 93K at optimum doping (p = 0.16) (Naqib et al., 2003).

The values of $E_g$ obtained from the analysis of the χ(T, p) data agrees well with the earlier reports (Naqib et al., 2003; Naqib et al., 2005a; Naqib et al., 2005b). A similar analysis of the χ(T, p) data for Ca-substituted Y123 also yields identical values for $E_g(p)$ once the p-independent Ca induced Curie term is subtracted. $E_g(p)$ values extracted from the fits are shown in Fig. 10 together with the $T_c(p)$ for pure Y123. We have obtained similar electronic phase diagrams using the onset of resistive downturn due to PG for different families of hole doped copper oxide superconductors (Naqib et al., 2003; Naqib et al., 2005a; Naqib et al., 2005b). The features are quite generic and agree very well qualitatively with Fig. 10.

**4. Discussion**

The $\rho_{ab}$ of copper oxide superconductors is the electrical resistivity measured parallel to the $CuO_2$ planes. It is one of the most revealing transport properties of these materials because it reflects how strongly correlated electrons move, scatter, and reorganize as temperature and doping change. The central physical idea is that in-plane transport is dominated by electronic motion within the $CuO_2$ planes, where strong electron–electron correlations of doped Mott system, antiferromagnetic spin fluctuations, and the PG profoundly modify the conventional metallic behavior.

The most remarkable feature of in-plane resistivity is its T-linear behavior over a very extended range of temperature encompassing 400 K to the onset of strong SC fluctuations when p ~ 0.19 (Figs. 1a and 1b). This is a hall mark of unconventional electronic ground state (Cooper et al., 2009; Loram et al., 2001; Tallon et al., 1999; Panagopoulos et al., 2002). For ordinary Fermi liquids (FLs), the electron-electron scattering leads to a $T^2$ dependence at low temperature while at higher temperatures comparable to the Debye temperature, a T-linear trend is expected (Ashcroft and Mermin, 1976). PG induced downturn in $\rho_{ab}(T)$ and this anomalous T-linear charge transport in cuprates have profound impact on our understanding of these strongly correlated electronic materials.

Several theoretical proposals attempt to explain the T-linear resistivity and its relevance to QP dynamics related directly to superconductivity and the PG (Cooper et al., 2009). Marginal Fermi liquid (MFL) (Varma et al., 1989) theory and nearly antiferromagnetic Fermi liquid (NAFL) (Millis et al., 1990; Millis, 1992) theory are two influential approaches to understand the anomalous normal-state properties of high-$T_c$ cuprate superconductors, particularly their unusual electrical resistivity, photoemission spectra, magnetic response, and low-energy electronic excitations. Both these theories attempt to explain why the normal state of cuprates often departs from the conventional Landau Fermi-liquid picture. However, they emphasize different physical mechanisms. Within MFL electrons interact with a phenomenological spectrum of low-energy electronic fluctuations that produces a scattering rate approximately linear in energy or temperature. While for NAFL, electrons interact strongly with collective spin fluctuations associated with an antiferromagnetic instability, particularly close to the wave vector **Q** = (π/a, π/a) in the Brillouin zone of a square-lattice $CuO_2$ plane.

The central distinction is that MFL describes the form of the electronic scattering spectrum, whereas NAFL identifies a more specific microscopic origin for strong scattering: nearly critical antiferromagnetic spin fluctuations which arises naturally from the proximity of the hole doped cuprates to its parent antiferromagnetic insulating phase. Both MFL and NAFL approaches predict low temperature T-linear resistivity when the PG vanishes. One particular advantage of NAFL formalism is that the proposed antiferromagnetic spin fluctuations can generate an effective Cooper pairing channel. Moreover, NAFL theory has also been used to explain aspects of the PG of cuprates through strong antiferromagnetic correlations and precursor spin-density-wave physics. In this scenario, sufficiently strong scattering between momentum states **k** and **k** + **Q** can reconstruct or suppress portions of the electronic spectrum, especially near the hot spots (π/a, 0) of the Brillouin zone.

A quantum critical point (QCP) in the electronic phase diagram is another possibility leading to a T-linear resistivity for a particular critical hole concentration, e.g., p ~ 0.19 (Cooper et al., 2009; Loram et al., 2001; Tallon et al., 1999; Panagopoulos et al., 2002). The key mechanism here is the interaction between conduction electrons and critical collective fluctuations, such as spin, charge, or nematic fluctuations. Close to the QCP, the QP scattering rate becomes approximately:

$$\frac{1}{\tau} = \alpha \frac{k_B T}{h} \qquad (4)$$

Here, τ is the transport relaxation time, $k_B$ is Boltzmann constant, h is Planck's constant and α is a dimensionless coefficient. This relation describes a characteristic Planckian scattering scale (Legros et al., 2019). This expression gives the physical basis of the T-linear resistivity of many strange metals. However, it is important to keep in mind that the Drude derivation is a useful interpretation rather than a complete microscopic theory of all quantum-critical electronic systems. The underlying transport mechanism remains an active research topic.

The same critical fluctuations that disrupt ordinary Fermi-liquid transport may, in some systems, provide an effective interaction that helps bind electrons into Cooper pairs. Whether this mechanism is dominant depends on the material and the type of critical fluctuations. Till now, the nature of critical fluctuations and symmetries related to quantum phase transition in hole doped copper oxide superconductors remain unidentified.

The discussion so far focuses on in-plane resistivity. The out-of-plane resistivity responds more significantly to the presence of PG and can be used to construct the electronic phase diagram of hole doped cuprates (Naqib et al., 2011). In tetragonal cuprate models, $\rho_c$ is controlled by the interlayer hopping amplitude, which can be approximated as (Naqib et al., 2011):

$$t = t_0[\cos(k_x a) - \cos(k_y a)]^2 \qquad (5)$$

where a is the in-plane lattice constant and $t_0$ sets the hopping scale. This expression has two important properties: (i) near the nodal points, where $k_x \sim k_y$, the hopping amplitude is strongly suppressed and (ii) close to the anti-nodal points, around ($\pi$/a, 0) and (0, $\pi$/a), the hopping amplitude is large. Therefore, c-axis charge transport disproportionately samples the anti-nodal electronic states in the momentum space. Hence, the tunneling model links the highly anisotropic structure of cuprates to their unusual electrical transport. Weak interlayer coupling makes c-axis conduction sensitive to the availability of electronic states in neighboring $CuO_2$ planes. The PG removes QP spectral weight particularly strongly around the anti-nodal region of the Brillouin zone. Thus, as PG suppresses the anti-nodal spectral weight, interlayer conductivity falls sharply and the c-axis resistivity rises strongly on cooling, even though in-plane transport remains metallic.

Uniform magnetic susceptibility is a bulk measure closely related to the thermodynamic parameters in hole doped copper oxide superconductors. Perhaps the most direct bulk probe of the EDOS is the electronic heat capacity measurements. The electronic specific heat coefficient, $\gamma_{el}(T)$, for Fermi liquids gives a direct measure of N($\varepsilon$) around the Fermi level. The zero magnetic field electronic entropy, $S_{el}(T)$, obtained from $\int\gamma_{el}(T)dT$ shows a striking qualitative and quantitative resemblance with the $\chi(T)T$ data for different families of cuprate superconductors including La214 and Y123 (Cooper and Loram, 1996; Loram et al., 1998a). This demonstrates that $\chi(T)$ probes the same underlying EDOS as probed by the heat capacity measurements. The electronic entropy measures the number of thermal excitations over the entire electronic spectrum (spin and charge over all wave vectors) whilst $\chi(T)T$ provides complementary information for the **q** = 0 spin spectrum. In fact, $\chi_{spin}(T)k_BT/\mu_B^2$ gives a measure of the number of thermally excited spin excitations at a particular temperature T (Loram et al., 1998a). Equation (1) relates electronic QP DOS to the Pauli spin susceptibility. These QPs carry both charge and spin. These also indicate that the PG observed in the spin sector is the same as that revealed when thermally excited QPs are probed. Such findings are hard to reconcile with scenarios where spin-charge separations are invoked to describe the normal state properties of cuprates (Loram et al., 1998a). As far as the Wilson ratio, $R = S_{el}/\chi T$, is concerned, its value is found to be quite close to that for ordinary Fermi liquids in cuprates over a wide range of hole concentration extending from the UD to the OD regions. This, at first sight, is surprising considering the wide body of evidences emphasizing the strongly correlated nature of the electronic ground state in cuprates (Loram et al., 1998a). The possible reason might be that the nodal excitations behave like Fermi liquids even in the presence of strong correlations. These nodal QPs should dominate bulk properties like heat capacity and static magnetic susceptibility. From the analysis of the heat capacity data, it was found that an electronic states non-conserving triangular gap pinned at the Fermi level describes the experimental $S_{el}(T)$ and $\gamma_{el}(T)$ quite satisfactorily (Loram et al., 1998a). Nearly parallel downward shifts of the $\chi(T)T$ data shown in Figs. 7 – 8, with no sign of their recovery at higher temperatures for cuprates with $p < 0.19$ are definitive signatures of loss of QP spectral weight due to PG. This states non-conserving nature of the PG is important because in conventional correlation driven spectral gaps lost states at low energies are recovered at high energies – the spectral weight is redistributed but not lost permanently.

Reconciling the PG phenomena observed across ARPES with its unique momentum resolution, uniform magnetic susceptibility ($\chi$), and temperature-dependent resistivity ($\rho$) is one of the central challenges in understanding the normal and superconducting state properties of cuprate high-$T_c$ compounds. While each probe measures a different physical quantity - single-particle spectral weight within the Brillouin zone, bulk spin susceptibility, and transport scattering rates, respectively - they collectively map out the multi-faceted energy scales and momentum-space anisotropies of the PG state below the characteristic temperature $T^*$.

ARPES provides momentum-resolved insight into the single-particle excitation spectrum. Below $T^*$, a gap appears sharply at the anti-nodal regions [($\pi$, 0) and (0, $\pi$)], while low-energy spectral weight persists along the nodal directions [$k = (\pm k_F, \pm k_F)$], creating the so-called Fermi arcs (LeBoeuf et al., 2007). Here, $k_F$ denotes the Fermi wave vector. The anti-nodal pseudogap manifests as a suppression of coherent QP peaks, signaling that electrons at the Brillouin zone boundaries lose their sharp quasiparticle integrity.

The drop in uniform susceptibility indicates a reduction in the density of states available for low-energy spin excitations at the Fermi level. Because the anti-nodal regions contribute heavily to the density of states in the cuprate d-wave band structure, the gapping of these exact anti-nodal regions observed in ARPES directly accounts for the loss of spin susceptibility resulting in the downturn in $\chi(T)$ in the normal state. This suppression in $\chi(T)$, sometimes has been interpreted as local pre-formed spin-singlet pairing or a precursor to singlet correlations, where phase incoherent paired electrons lose their spin response to an external magnetic field.

As seen in the preceding section, in the normal state well above $T_c$, hole doped cuprates exhibit the famous linear-in-T resistivity, a decisive hallmark of the "strange metal" regime. The emergence of the pseudogap (as seen in ARPES) partially gapping the Fermi surface removes portions of the Fermi surface available for carrier scattering. Furthermore, the suppression of low-energy spin fluctuations (tracked by the uniform magnetic susceptibility) reduces the scattering rate of quasiparticles due to antiferromagnetic spin fluctuations. Lower scattering phase space directly translates to a drop in electrical resistivity.

## 5. Conclusions

Considering finer details, existing theories addressing the origin of the PG can be grouped into four different classes (Li and She, 2022), namely, a precursor to a low energy ordered state [e.g., superconductivity (Kordyuk, 2015; Li, 2021) or spin density wave (Sedrakyan and Chubukov, 2010; Schmalian et al., 1999)], a QP spectral gap due to band folding induced by spin density wave (SDW) or charge density wave (CDW) order, a hybrid spectral gap induced by intertwined orders [(Schmalian et al., 1999; Comin and Damascelli, 2016) e.g., spin/charge stripe (Fradkin et al., 2015) and pair-density waves (Tranquada et al., 1995; Agterberg et al., 2020)], and a QP spectral feature originating from strong correlations [e.g., residual antiferromagnetic fluctuations expected in doped antiferromagnets (Arovas et al., 2022; Chakravarty et al., 2001)] without breaking of any symmetry associated with a phase transition. Although these theories have qualitatively described some pseudogap properties, none of them can precisely describe the whole pseudogap phase and all essential experimental manifestations; thus, there is no wide consensus on the origin of the pseudogap yet (Li and She, 2022; Tallon and Loram, 2001; Timusk and Statt, 1999). For instance, though the precursor Cooper pairing to phase coherent superconductivity could explain the $T_c(p)$ dome, it predicts that $T^*(p)$ merges with the superconducting dome on the overdoped side, contradicting the experimental observation that PG appears to vanish quite abruptly at a critical doping of $p^* \sim 0.19$ (Cooper and Loram, 1996; Loram et al., 1998a; Naqib et al., 2003; Naqib et al., 2005a; Naqib et al., 2005b). On the other hand, numerical solutions of the Hubbard model tends to obtain a correct $T^*(p)$ line with reasonable set of parameters, but it is difficult to accurately predict the nature of phase boundaries and the competition between density wave orders and uniform phases at a finite temperature. Furthermore, intertwined electronic orders involving both particle-particle (i.e., superconductivity) and particle-hole (e.g., different density orders) interaction channels provide a comprehensive description of the ARPES results (Kordyuk, 2015; Varma, 2019), but the stability and the precise nature of the emergent symmetry remains questionable in case of hole doped copper oxide superconductors.

The origin, hole content dependence, and energy/temperature evolution of PG in cuprate superconductors therefore remains a central puzzle at the intersection of strong correlation physics, unconventional superconductivity, and quantum criticality. The remarkable asymmetry between hole- and electron-doped cuprates has provided crucial experimental constraints that have driven theoretical progress (Armitage et al., 2010; Duffy et al., 2026) but a definitive answer regarding the origin and evolution of the PG in hole doped systems remains unclear.

For hole-doped cuprates, the evidence suggests that competing orders, particularly charge-density-wave and stripe correlations, play a significant role in PG formation, though their relationship to superconducting preformation remains debated.

The bulk transport and magnetic susceptibility data presented in this work strongly supports non-superconducting origin of the PG. The strong signatures are the facts that (i) resistivity scaling fails for overdoped samples and (ii) the $T^*(p)$ line falls below the $T_c(p)$ dome and terminates at a critical hole concentration, $p^* \sim 0.19$. These observations cannot be accommodated within scenarios where PG is derived from phase incoherent precursor Cooper pairings. The precise nature of the quantum phase at the critical hole content is not understood at this moment.

Our findings also suggest that PG competes with superconductivity via the removal of QP spectral weight. This explains the drastic reduction of SC condensation energy, ground state superfluid density, and the critical current density in high-$T_c$ cuprates in the UD region affected by the presence and magnitude of the PG (Loram et al., 1998a; Naqib and Semwal, 2005; Naqib and Islam, 2019; Naqib and Islam, 2023; Ahmad and Naqib, 2020).

Ultimately, due to the rich and complex electronic phase diagram of hole doped cuprates, the PG might be a manifestation of multiple intertwined phenomena - competing orders, effect of disorders (Tallon et al., 2006; Naqib et al., 2005a; Naqib, 2007; Islam and Naqib, 2008; Islam and Naqib, 2010; Islam and Naqib, 2018; Naqib and Islam, 2014), precursor fluctuations, and intrinsic correlation effects - whose relative importance varies with doping and energy/temperature (Naqib and Islam, 2015; Borna et al., 2022).

Experimental refinements coupled with advances in theoretical and computational approaches continue to provide more insights into this long-standing PG puzzle, and this may, in turn, reveal the essential physics of high-temperature superconductivity itself.

## Acknowledgements

The authors acknowledge long standing research collaborations with Emeritus Professor J. R. Cooper, Quantum Matter Group, Cavendish Laboratory, University of Cambridge, UK, Dr. J. W. Loram, Quantum Matter Group, Cavendish Laboratory, University of Cambridge, UK, and Professor J. L. Tallon, Robinson Research Institute at Victoria University of Wellington, New Zealand for their scholarly insights which helped the attempts to understand the pseudogap phenomenon in copper oxide superconductors. We dedicate this work to the fond memories of late Professor M. Shamsher Ali, who inspired generations of physicists in Bangladesh. The authors also thank Dr. Nazir Ahmad for his help in managing the references.

**Conflict of Interest**

The authors declare no conflict of interest

**Data availability statement**

The datasets used in the manuscript will be available upon reasonable request to the corresponding author.

**CRediT authorship contribution statement**

**Saleh Hasan Naqib:** Conceptualization, Experimentation, Formal analysis, Writing-original draft, review & editing; **Raihana Shams Islam:** Validation, Writing-Reviewing and Editing.

**Declaration of Generative AI and AI-assisted technologies in the writing process**

During the preparation of this manuscript, the authors used ChatGPT (OpenAI) in order to improve the language, grammar, readability, and organization of the manuscript. After using this tool, the authors reviewed and edited the content as needed and take full responsibility for the content of the published article.